\documentclass[10pt, preprint2]{aastex}

\begin{document}

\title{On The Depolarization Asymmetry Seen in Giant Radio Lobes.}

\author{M.B. Bell\altaffilmark{1} and S.P. Comeau\altaffilmark{1}}

\altaffiltext{1}{National Research Council of Canada, 100 Sussex Drive, Ottawa,
ON, Canada K1A 0R6;
morley.bell@nrc-cnrc.gc.ca}

\begin{abstract}

The depolarization asymmetry seen in double-lobed radio sources, referred to as the Laing-Garrington (L-G) effect where more rapid depolarization is seen in the lobe with no visible jet as the wavelength increases, can be explained either by internal differences between the two lobes, or by an external Faraday screen that lies in front of only the depolarized lobe. If the jet one-sidedness is due to relativistic beaming the depolarization asymmetry must be due to an intervening Faraday screen. If it is intrinsic the depolarization asymmetry must be related to internal differences in the lobes. For a random viewing angle distribution, which must be the case here where un-beamed lobe radiation dominates, jet one-sidedness is unrelated to viewing angle and therefore cannot be used either to estimate the viewing angle or to imply beaming. The outflow speed in the kpc jet is notoriously difficult to determine. However, although it has not yet been proven conclusively, we assume in this paper that the speed in the outer jet of several Fanaroff-Riley Class 1 (FRI) sources exhibiting the L-G effect is close to the 0.1c reported by several other investigators. For these sources we find that the jet one-sidedness cannot be explained by beaming and therefore must be intrinsic. In these FRI sources the L-G effect must be due to differences that originate inside the lobes themselves, with the outer regions of the relevant lobe acting as a Faraday screen. Although it is not known if the flow in the outer jets of FRII sources also slows to this speed it is suggested that the explanation of the L-G effect is likely to be the same in both types. This argument is strengthened by the recent evidence that FRII galaxies have very large viewing angles, which in turn implies that the L-G model cannot work regardless of the jet velocity. It may therefore be too soon to completely rule out internal depolarization in the lobes as the true explanation for the L-G effect.

\end{abstract}

\keywords{galaxies: active --- galaxies: distances and redshifts --- galaxies: jets --- quasars: general}

\section{Introduction}

In double-lobed radio sources with single-sided jets more rapid depolarization has been observed with increasing wavelength in the lobe on the side with no jet (the Laing-Garrington effect). Important in the explanation of this effect is the question of whether the jet one-sidedness is \em intrinsic \em or due to \em relativistic beaming. \em If it is the latter, and both lobes are still being actively fed, no significant internal differences would be expected in the lobes and the depolarization asymmetry must then originate in an external Faraday screen. Furthermore, in this model the lobe on the side with no jet will be receding and its radiation will be expected to be depolarized to a greater degree if it is viewed through more of a Faraday screen associated with a halo surrounding the parent galaxy \citep{gar88,lai88,gar91}. This model works best if the jet viewing angles are small so that the path length through the screen to the more distant, more highly depolarized lobe, will be a maximum. Small jet viewing angles can also be a natural result if relativistic beaming plays a significant role.

On the other hand, if the jet one-sidedness is intrinsic there is an equal chance that the lobe on the side containing the visible jet is moving away from us, which means that the intervening Faraday screen model no longer works and the depolarization asymmetry must then originate inside the lobe itself, with the outer regions of the lobe acting as a Faraday screen. The fact that no jet is seen on one side implies that the jet is either much weaker or turned off completely. In this case high energy particles are no longer being pumped into its associated radio lobe at the same rate and the conditions inside that lobe will be expected to be different than in the lobe that is still being actively fed, which would be required if the depolarization is internal. Evidence that the lobe dissipates with time when it is no longer being fed is easily seen in the trailing edges of the lobes in many of the double-lobed sources, such as 3C66B or 3C 296 \citep{bri06}. It might also be expected that if one of the lobes is no longer being fed its outward motion will slow down, with the result that it will be located closer to the central compact object as is often seen. However, this observation can also be explained in the beaming model by differences in light travel time.

Before proceeding further it is important to note that there is a third type of Faraday screen that can add some confusion to this analysis. It has been demonstrated that there can sometimes be weak Faraday screens present that cover the entire source \citep{per84,ode86,dre87,lai87} and these are most easily seen in those sources with two jets \citep[see for example]{bes98,goo04} where both lobes are still being fed equally and the asymmetric depolarizing component we are discussing here is either very low or not present. These screens can be due either to extended material in the inner regions of clusters, material in our own Galaxy, or intergalactic material, but they should not be confused with the asymmetric Faraday screen proposed in the L-G sources which covers only one lobe.

\section{Previous Work}
 
Because it was assumed by \citet{gar91} that the jet one-sidedness was likely due to relativistic beaming it was concluded that the Laing-Garrington effect is more likely to originate in an intervening Faraday screen that covers the receding lobe and this explanation has endured. In fact, in most investigations since then it has also been assumed without proof that the one-sidedness is due to relativistic beaming. As an example, \citet{mor97} set out to see if the L-G effect was present in low-luminosity radio galaxies. It was thought that the jets in these objects would be less likely to be relativistic and they argued that a positive detection for the L-G effect would prove that they were. However, that argument is only valid if the relativistic beaming model is the correct explanation for the jet one-sidedness, or if it could somehow rule out the internal explanation. Their analysis included two samples; one with single-sided jets and one with two-sided jets.
Only the sample with one-sided jets showed a significant L-G effect. However, although their result is consistent with the L-G explanation if the one-sidedness is due to beaming, it is also consistent with the internal depolarization explanation if the one-sidedness is intrinsic. In the two-sided case both lobes are still being fed so no differences in the depolarization would be expected in either the beaming or intrinsic models. In the single-sided case the depolarization asymmetry can be explained in the intrinsic case by the fact that only one lobe is being fed.

These authors also found that the radio galaxies with strong radio cores showed a more pronounced asymmetry in depolarization. This is also what would be expected in the internal depolarization model where the depolarization is tied directly to the rate at which the lobe is being fed. It is thus apparent that, although their results may be consistent with the L-G explanation, they are also not inconsistent with the internal depolarization explanation.

If the jets are intrinsically one-sided, the lobes are being excited in different ways so different internal Faraday depths might be expected \citep{liu91}. These authors found differences in the spectra between the two lobes and suggested that a possible explanation might be due to differences in the physical conditions in the lobes. They suggested that possible differences might arise in the rates of deposition of energy from the nucleus, and this would certainly be the case if one jet had turned off. They concluded that there must be differences in the physical parameters of the two lobes which influence both spectrum and depolarization.

\begin{figure}[t]
\hspace{-1.0cm}
\vspace{-1.0cm}
\epsscale{0.9}
\includegraphics[width=9cm]{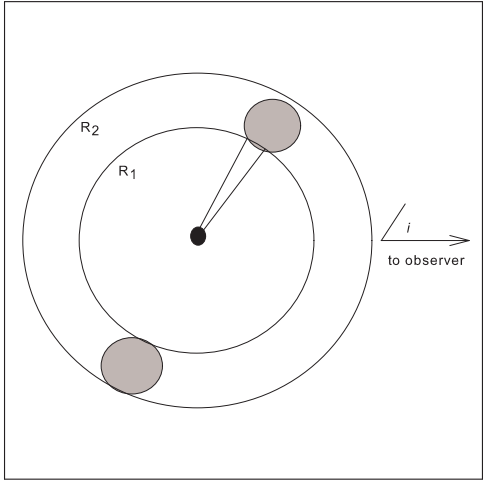}
\caption{{Source with two radio lobes viewed at the mean angle $i$ expected for random orientations, where $i$ is the angle between the jet and the line-of-sight as shown. See text for an explanation of R$_{1}$ and R$_{2}$. \label{fig1}}}
\end{figure}

The Laing-Garrington effect was also looked at closely by \citet{gop00}. They point out that if this effect is due to an intervening Faraday screen of the nature proposed by \citet{gar91}, the viability of the explanation demands that the Faraday screen somehow maintains a size that is within a factor of two of the total extent of the expanding radio source, which may be difficult to explain. In Fig 1 no difference in depolarization will be expected if the radius of the Faraday screen is R $<$ R$_{1}$. The biggest effect will occur for R = R$_{2}$, but even then the two path lengths will not differ significantly. The situation gets worse again for R $>$ R$_{2}$. For very large R, as would be the case for screens that cover the entire central regions of clusters the difference in path length to the two lobes would be negligible, as would be the effect. It seems unlikely that the halo would just happen to be the optimum size R = R$_{2}$. Furthermore, there is no solid evidence that the proposed Faraday screen (galactic halo) even extends out as far as the lobes, which can be up to 1500 kpc in projected linear size. There is clearly no evidence that it does in the map of 3C296 \citep{bri06} where the central elliptical galaxy, and its optically bright halo, are superimposed on top of the 20 cm radio map. Here both lobes lie well outside the optical halo of the galaxy with neither being viewed through it.

Because of these problems \citet{gop00} suggested that the superdisks seen in some radio galaxies might be an alternative to the Faraday screen envisioned by \citet{lai88,gar91} that would remove the need to postulate a magnetoionic corona around every high-z quasar whose dimensions somehow remain closely matched to the steadily growing radio source. However, the superdisks they discuss have so far only been seen in radio galaxies and whether or not they even exist in quasars is unclear. They do not appear to be common even in radio galaxies.

\section{Observing sources that have giant radio lobes with internal hotspots}

Although the flow in the jet may start out relativistically at parsec scales, there is much evidence from radio maps that the flow slows down considerably on kiloparsec (kpc) scales \citep{bic94,lai99,bri06}. It is these large kpc-scale jets that the L-G investigations have used to conclude that beaming is present. Unlike the pc-scale jet motion near the core that can easily be measured with the VLBA to milliarsec (mas) accuracy, no direct method has been found to measure the motions of material in the kpc-scale jets to mas accuracy. This is because in this case the VLA must be used and detecting mas motions with an arcsec beamwidth is next to impossible. Although \citet{wal88} have claimed to have made such a measurement in the kpc jet of 3C120, this measurement has not been confirmed. But it is well known that radio galaxies, especially the closer ones, do not show the highly relativistic motions in their pc-scale jets that are seen in many quasars \citep{kel04}(see their Fig 6).


The L-G explanation works best if the jet viewing angles are small so that the path difference to the two lobes is greatest. But there are several things that need to be kept in mind when dealing with sources with giant radio lobes. For these sources the dominant component of the flux density in the early surveys, such as the 3C, 4C, or Parkes surveys, that were carried out at low frequencies, more than 90 percent of the flux comes from the lobes and their hotspots \citep{bri84}. Because the lobes are separating from the core at relatively slow speeds from 0.01c to 0.03c \citep{ale87,cle07,ode09}, their radiation cannot be Doppler boosted. \em By definition, the L-G sources all fall into the lobe-dominant category. \em The percentage of un-beamed radiation is even higher if the core radiation is not beamed as has recently been claimed \citep{bel10}. When the dominant component of the radiation is not beamed there can be no selection effect that will preferentially pick up those sources with small jet viewing angles \citep{bel12}. It was also pointed out by \citet{cle07} that using low frequencies in source surveys, as was the case for the early source-finding surveys that found most of the radio-loud sources, provides a unique way of obtaining an \em orientation unbiased \em sample of AGNs. When there are no selection effects to preferentially pick up sources with specific viewing angles, \em all \em sources above the detection limit in this category (with giant lobes) will be detected. In this case the number distribution of their orientations will be random (sin$i$) as shown in Fig 2 of \citet{bel12}. Here, and throughout this paper, $i$ is the angle between the jet and the line-of-sight. For a random distribution, $50\%$ will have viewing angles above $60\arcdeg$. Less than 1 percent will have viewing angles below $8\arcdeg$. This means that very few of the detected sources will have the small viewing angles required for the L-G model to work effectively.

The situation may be different for core-dominant quasars. Most astronomers who study superluminal motion argue that in searches, Doppler boosting will preferentially pick up those sources with their jets pointing towards us \citep{kel04,ver94,lis97,lop12}, although some have claimed otherwise \citep{bar89, bel12}. However, whether or not this is the case for the core-dominant sources (quasars), these are not being examined here. Here we are looking at L-G sources. These are sources that by definition all have giant radio lobes and have Fanaroff-Riley classifications FRI and FRII \citep{fan74,sar12}. For the L-G radio galaxies being considered here, those sources with their jets pointed in our direction will not have been $preferentially$ selected in the original surveys.

What is also important to realize here is that when giant lobes are present the jet one-sidedness cannot be an indication of relativistic beaming since the jet component is too small to be able to change the distribution from a random one. This means that, \em essentially the same viewing angle distribution would be obtained regardless of whether the jet strength was intrinsic or beamed. This, in lobe-dominant radio galaxies, makes it very difficult for the L-G model to work even if the kpc jet is relativistic. \em

\section{Deceleration in kpc-scale jets}

Several recent investigations have been carried out to determine how the flow decelerates in the jet as it moves outward \citep{lai02a,lai02b,lai06,lai07}. It was concluded \citep{lai07} that jets in FRI radio galaxies, although initially relativistic with velocities from 0.8c to 0.9c at parsec scales, decelerate rapidly in the first few kpc to non-relativistic velocities between 0.4c and 0.1c beyond the flaring region. Since in most cases only about the first ten percent of the kpc jet was studied by these investigators, and it is obvious from the radio maps that most continue to decelerate, these values are likely to be upper limits. Furthermore, \citet{lai99} also found that beyond a few kpc the jet flow decelerates to 0.1c. It is therefore assumed here that, at least for FRI sources, when the entire kpc jet is considered the average outward speed (beyond $\sim10$ kpc) is likely to be no more than 0.1c. This would be completely consistent with the outward motions of the lobes which have been shown to be close to $0.02c\pm0.01c$ in both FRI and FRII sources \citep{ale87,cle07,ode09}. For these reasons we have assumed in this study that at least for the low luminosity, FRI galaxies studied here that the kpc flow speed is likely to be close to 0.1c in the outer jet.

\begin{figure}[t]
\hspace{-0.8cm}
\vspace{-1.0cm}
\epsscale{1.0}
\includegraphics[width=9cm]{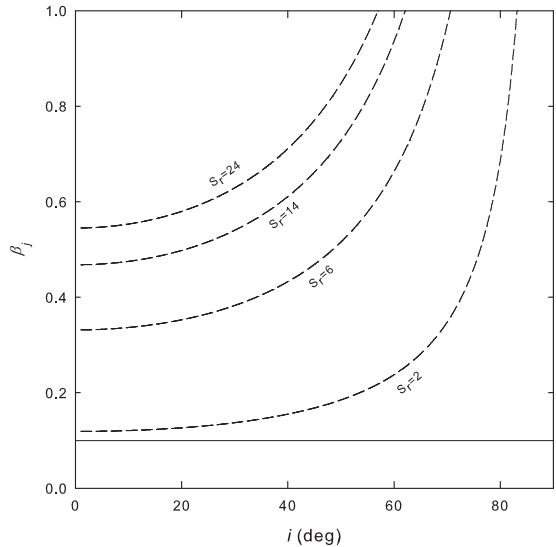}
\caption{{Jet velocity (in units of c) required to produce different S$_{r}$ values by beaming effects plotted as a function of jet viewing angle \em i. \em The horizontal line at $\beta_{j}$ = 0.1 represents the flow in the kpc jet found for FRI sources. Clearly at this flow rate even an asymmetry as low as S$_{r}$ = 2 cannot be completely explained by beaming even for small viewing angles. See text for further discussion. \label{fig2}}}
\end{figure}

\section{Is the jet asymmetry intrinsic or is it due to beaming?}

For a given jet asymmetry ratio (S$_{r}$ = S$_{j}$/S$_{cj}$) it is possible to calculate how the jet flow speed must increase as a function of viewing angle $i$ from the relation

 S$_{j}$/S$_{cj}$ = [(1+$\beta_{j}cosi$)/(1-$\beta_{j}cosi$)]$^{2-\alpha}$

Here S$_{j}$ and S$_{cj}$ are the fluxes in the jet and counterjet respectively, $\beta_{j}$ is the jet velocity in units of c, and $\alpha$ is the spectral index \citep{hoc10,urr95}. 

In Fig 2, $\beta_{j}$ is plotted against the viewing angle $i$ assuming several values for S$_{r}$. If the jet flow is assumed to be 0.1c as shown by the solid line in Fig 2, the maximum jet asymmetry that can be produced by beaming effects is less than S$_{r} = 2$, even for very small jet viewing angles. Although a kpc jet speed of 0.1c in these sources has not been proven conclusively, it has been reported by several investigators to slow to close to this value in the outer jet \citep{lai02a,lai02b,lai06,lai07}. For a jet flow speed of 0.9c, an asymmetry of S$_{r}$ = 24 cannot be produced by beaming effects if the jet viewing angle is greater than $\sim60\arcdeg$.

But as discussed above, although the flow may start out close to 0.9c when it first spews out of the core, it slows down significantly in the outer (kpc) jet where for FRI sources it may have mean values closer to 0.1c. This slows further to speeds near 0.02c $\pm0.01c$ in the lobes \citep{ale87,cle07,ode09}.

\begin{figure}[t]
\hspace{-0.8cm}
\vspace{-1.5cm}
\epsscale{1.0}
\includegraphics[width=9cm]{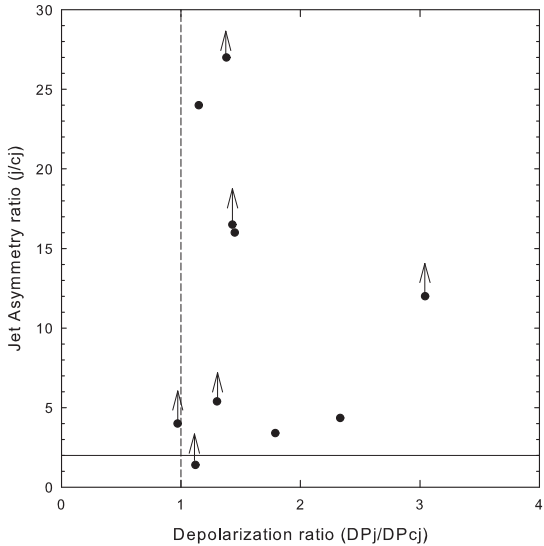}
\caption{{Jet asymmetry ratio, S$_{r}$, versus depolarization ratio for low-luminosity sources from \citet{mor97}. \label{fig3}}}
\end{figure}

Variations in the spectral index of $\pm$0.3 produced small shifts in the dashed curves in Fig 2 but these were nowhere near big enough to affect these conclusions. Since the outer-jet flow value of 0.1c that we have assumed here has been obtained from low-luminosity FRI sources it is necessary to use these sources for this investigation.

In Fig 3 the kpc jet asymmetry ratio, found by \citet{lai99} for the single-sided low-luminosity FRI sources studied by \citet{mor97}, is plotted versus their corresponding jet depolarization ratio. The up arrows indicate those sources with lower limits. Sources that show the L-G effect must lie to the right of the vertical dashed line and as reported by \citet{mor97} these one-sided sources all appear to show this effect. However, if the outer jet flow in low-luminosity sources is assumed to slow to $\sim$0.1c beyond a few kpc, as has been shown to be the case by several investigators, from Fig 2 relativistic beaming can only explain those jet asymmetries that fall below the solid horizontal line in Fig 3. All of the asymmetry that lies above this line must then be intrinsic. From Fig 3 it can be seen that most of the jet asymmetry in these low-luminosity sources must be intrinsic if our assumption of an outer jet speed near 0.1c is correct. 

These results argue strongly that when the asymmetry in the kpc-scale jet is large it must be mostly intrinsic and not due to relativistic beaming. This in turn implies that the L-G effect in these sources originates inside the lobe and not in an external Faraday screen. This result also allows for much larger viewing angles, which is more consistent with that predicted for a random distribution.

\begin{figure}[t]
\hspace{-0.8cm}
\vspace{-1.5cm}
\epsscale{1.0}
\includegraphics[width=8.5cm]{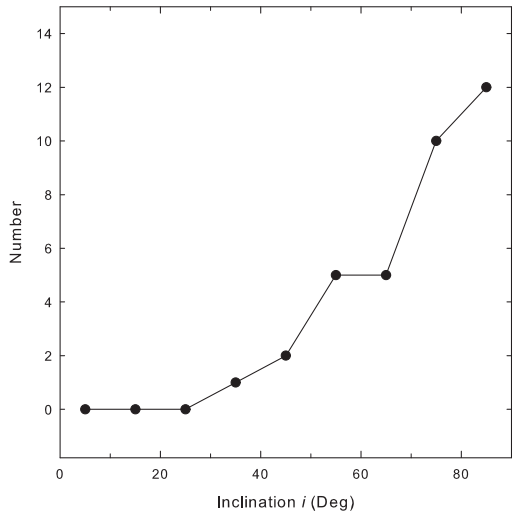}
\caption{{Number of FRII radio galaxies per 10 deg inclination interval from \citet{dro12}, where $i$ is the angle between the jet and the line-of-sight.
\label{fig4}}}
\end{figure}

Although the outflow speed in the jets of FRII sources is not as accurately known it seems unlikely that the L-G explanation would differ between the two types, and this would require that it be similar to that of the FRI sources. However, there are other reasons why this is likely to be the case. First, it has already been found that the outward motion of the lobes in FRII sources is similar to that seen in FRI sources \citep{ode09}, Second, it has already been found that the viewing angles of FRII galaxies may be even larger than those of FRI galaxies. Recently \citet{dro12} have estimated the jet inclination angles $i$ in several FRII radio galaxies selected from a larger sample to cover a redshift range from 1 to 5.2. They estimate the inclination angle using the core dominance factor R, which measures the relative strengths of the core and inner jet flux to the extended flux component. This should work well for the lobe-dominant sources where the core flux can be referenced against a much stronger extended component. Whether it works as well for core-dominant sources (quasars) is questionable since the extended flux component is much weaker and may vary significantly from source to source. Their results show that most of the sources in their sample that are classified as FRII sources by NED have jet inclinations relative to the line-of-sight near 85 degrees. This is as large, or larger, than expected for FRI viewing angles, but it should not be unexpected since these sources, like the L-G sources being discussed here, all have giant lobes containing most of the flux. Fig 4 shows the jet viewing angle $i$ distribution for the galaxies as measured by \citet{dro12} and listed in their Table 5. Most of the sources have jet viewing angles near 80 degrees, which means that their giant lobes will also be close to the plane of the sky. These viewing angles are much too large for the L-G model to work effectively, which is completely consistent with what we found above for FRI galaxies.

\section{Other evidence that the jet one-sidedness is intrinsic}

Although an intrinsic explanation for one-sidedness in jets might have seemed unlikely 20 years ago, it has recently been shown to be the case in M87 where a special attempt was made to show that the jet asymmetry observed could be explained by relativistic motion \citep{kov07}. It was found that at 15 GHz the material producing the asymmetric radiation was not moving relativistically, and it must be concluded that here, too, the asymmetry is likely intrinsic. It is unfortunate that direct measurements of the motions in kpc jets are almost impossible to make since they require milliarcsec accuracy but must work with the arcsec resolution provided by the VLA.

\section{Conclusions}

Although the speed in the outer jet is difficult to measure we have assumed here that the flow in the low-luminosity FRI sources we studied slows to near 0.1c in the outer jet, as was reported previously by several other investigators. 
The main conclusions obtained here can then be summarized as follows. For the low-luminosity sources with highly asymmetric jets and which show the L-G effect, if their kpc-scale jet flow decelerates as quickly as has been claimed by several previous investigators the jet asymmetry cannot be produced by beaming. The asymmetry must then be intrinsic. If the jet one-sidedness is intrinsic there is an equal chance that the strongest jet is directed away from us. This means that the currently accepted L-G model cannot work because the proposed Faraday screen would then lie behind the more rapidly depolarized lobe associated with the weaker jet. This means that the depolarization must occur inside the relevant lobe with the outer regions of the lobe acting as a Faraday screen, and the currently accepted model is then unlikely to be correct. Although the jet flow in FRII sources is less accurately known, it seems likely that the explanation for the depolarization in these sources would be the same. Furthermore, if the viewing angles of FRII sources are very large, as recent results seem to show is the case, the L-G model cannot work because the lobes of these sources would be close to the plane of the sky. It may be too soon then to completely rule out internal depolarization in the lobes as the true explanation for the L-G effect.


\end{document}